\documentclass[prd,aps,prd,twocolumn,showpacs,preprintnumbers,nofootinbib,amsmath,amssymb, superscriptaddress]{revtex4-1}
\usepackage{graphicx}
\usepackage{color}
\usepackage{amsmath}
\usepackage{lineno}
\usepackage[normalem]{ulem}
\usepackage{hyperref}
\hypersetup{
    colorlinks=true,
    linkcolor=blue,
    filecolor=magenta,      
    urlcolor=cyan,
    citecolor=blue,
}
 
\usepackage{mwe}
\usepackage{subfig}

\begin{document}
\title{Enhancing the sensitivity of transient
gravitational wave searches with Gaussian Mixture Models} 
\author{V. Gayathri}
\affiliation{Department of Physics, Indian Institute of Technology Bombay, Powai, Mumbai 400 076, India}
\affiliation{Department of Physics, University of Florida, PO Box 118440, Gainesville, FL 32611-8440, USA}
\author{Dixeena Lopez}
\affiliation{Department of Physics, Indian Institute of Technology Bombay, Powai, Mumbai 400 076, India}
\affiliation{Physik-Institut, University of Zurich, Winterthurerstrasse 190, 8057 Zurich, Switzerland}
\author{R. S. Pranjal}
\affiliation{Department of Physics, Indian Institute of Technology Bombay, Powai, Mumbai 400 076, India}
\author{Ik Siong Heng}
\affiliation{SUPA, School of Physics and Astronomy, University of Glasgow, Glasgow G12 8QQ, United Kingdom}
\author{Archana Pai}
\affiliation{Department of Physics, Indian Institute of Technology Bombay, Powai, Mumbai 400 076, India}
\author{Chris Messenger}
\affiliation{SUPA, School of Physics and Astronomy, University of Glasgow, Glasgow G12 8QQ, United Kingdom}

\begin{abstract}
Identifying the presence of a gravitational wave transient buried in
non-stationary, non-Gaussian noise which can often contain spurious noise
transients ({\it glitches}) is a very challenging task. For a given data set,
transient gravitational wave searches produce a corresponding list of {\it
triggers} that indicate the possible presence of a gravitational wave signal.
These triggers are often the result of glitches mimicking gravitational wave
signal characteristics. To distinguish glitches from genuine gravitational wave
signals, search algorithms estimate a range of trigger attributes, with
thresholds applied to these trigger properties to separate signal from noise.
Here, we present the use of Gaussian mixture models, a supervised machine
learning approach, as a means of modelling the multi-dimensional trigger
attribute space. We demonstrate this approach by applying it to triggers from
the coherent Waveburst search for generic bursts in LIGO O1 data. By building
Gaussian mixture models for the signal and background noise attribute spaces,
we show that we can significantly improve the sensitivity of the coherent
Waveburst search and strongly suppress the impact of glitches and background
noise, without the use of multiple search bins as employed by the original O1
search. We show that the detection probability is enhanced by a factor of 10, 
leading enhanced statistical significance for gravitational wave signals such as GW150914.
\end{abstract}

\maketitle

\section{Introduction}

The era of gravitational-wave (GW) astronomy began with the first direct
detection of the GW signal observed on September 14,
2015~\cite{GW170814-DETECTION} by the Advanced
LIGO~\cite{TheLIGOScientific:2014jea} and Virgo~\cite{Virgo-2015Acernese}
detectors. So far, in the proceeding observing runs, the LIGO Scientific and
Virgo Collaboration have fourteen confirmed GW
detections~\cite{GW150914-DETECTION,GW151226-DETECTION,GW170104-DETECTION,GW170608-DETECTION,GW170814-DETECTION,TheLIGOScientific:2016pea,O1-O2-catalog-paper,
GW170817-DETECTION, GW190425-DETECTION}, with data from several further
candidate compact binary merger candidates still being analyzed. In the next
few years, the Advanced LIGO and Virgo detectors will improve in sensitivity,
and additional GW detectors such as KAGRA \cite{Kagra-design}, and LIGO-India
\cite{LIGO-India} will join the network. The sensitivity improvement of the GW
detector network will lead to more detections of GWs from different types of
source~\cite{Davis:2020nyf}.

With improved detector perfromance, GW detectors are becoming more sensitive to
other disturbances culminating in the observation of a large number of
non-Gaussian noise transients known as {\it
glitches}~\cite{TheLIGOScientific:2016zmo}. They originate due to complex
instrumental and environmental effects within GW detectors. Often, they produce
high signal-to-noise ratio (SNR) triggers in GW searches, increasing false
alarm rates and reducing search sensitivity. 
Data quality investigations~\cite{Nuttall_2015} that focus on the correlation
between instrument or environmental effects with GW data as well as veto
techniques have helped to eliminate some known glitch classes. However, a large
number of unknown noisy transients persist. Different searches explore
different types of glitch rejection
methods~\cite{Allen:2004gu,Nitz:2017lco,Canton:2013joa,Gayathri:2019omo} to
improve search sensitivity. Though important, these methods also increase the
computational burden of existing GW detection algorithms.

The volume of literature on the application of machine learning methods in GW
astrophsyics is rapdily growing. This includes techniques to improve signal
detection, parameter estimation of transient signals, noise removal techniques,
modeling GW signals as well as source population
inference~\cite{BAHAADINI2018172,PhysRevD.95.104059,Colgan:2019lyo,Cuoco:2020ogp,PhysRevD.101.042003,
George:2017pmj,Gebhard:2019ldz,Gabbard:2019rde,Cuoco:2020ogp}.

In this work, we propose a supervised machine learning method using {\it
Gaussian mixture models} (GMMs) for use in signal detection where we construct
two distinct models for noise glitches and astrophysical GW signals. Rather
than developing this approach using time-series data, we instead apply it after
the coherent Waveburst (cWB) algorithm has identified {\it triggers} --
interesting time instances which could be either potential GW signals or the
noisy transients. At present, the cWB search algorithm has multiple output
attributes and based on the location of the trigger attributes in this
multi-dimentional space, the trigger is classified as a GW event or noisy
event. We propose an alternative to the existing thresholding procedure on the
multi-dimentional attribute space by folding in the GMM naturally in the
detection problem under the likelihood ratio approach.

The paper is organized as follows : In Sec.~\ref{GMM}, we discuss the Gaussian
mixture modelling of the multi-modal data set. In Sec.~\ref{Sec:GMMLLR} we
discuss the use of the GMM in the construction of a log-likelihood based
detection statistic. In Sec.~\ref{burst} we assess the detection performance of the 
proposed GMM based detection method with respect to the generic burst algorithm
in an all-sky short-duration GW burst set-up. We apply the algorithm for
the coincident events from the first observing run of advanced LIGO detectors.
Finally, in Sec.~\ref{conclusion}, we discuss our conclusions.

\section{Gaussian mixture model}
%
\label{GMM}
Gaussian mixture models (GMMs) are probabilistic models which use
uni-modal Gaussian distributions to represent a multi-modal data set. Under
the GMM approach, a given data set is modelled as a weighted sum of a
collection of Gaussians. 

Let the data vector ${\bf x}$ be characterized by $d$ number of attributes. We
refer to these data as a $d$-dimensional data vector. The corresponding GMM of the data
consists of a superposition of $K$ Gaussian distributions
and is given by 

\begin{align}
    p({\bf x}) = \sum_{j=1}^{K}w_j \mathcal{N}({\bf x}|\boldsymbol{\mu_j}, \Sigma_j),
\end{align}
where, $\mathcal{N} ( {\bf x}|\boldsymbol{\mu_j} , \Sigma_j )$ is a multinomial
Gaussian distribution with $d$-dimensional mean vector $\boldsymbol{\mu_j}$ and $d
\times d$ co-variance matrix $\Sigma_j$ and is written as
\begin{align}
   \mathcal{ N} ( {\bf x}|\boldsymbol{\mu_j} , \Sigma_j ) = \frac { \exp \left[ -\frac { 1
} { 2 } ({\bf x} - \boldsymbol{\mu_j} ) ^ { T } \Sigma_j^{ - 1 } ( {\bf x} - \boldsymbol{\mu_j})
\right] } { ( 2 \pi ) ^ { d / 2 } |\Sigma_j| ^ { 1 / 2 } }. 
\end{align}
The parameter $w_j$ is the weight corresponding to each Gaussian component
normalised such that $\sum_j w_j=1$.

When the data is believed to contain two distinct populations but where each
population has complex but distinct structure within the attribute space, GMMs
can be used to model each population separately. These models can be then
incorporated into a likelihood-ratio test statistic which can be used to
identify events from either population. This is classed as an supervised
machine learning algorithm. 

\section{Detection method using GMM}
\label{Sec:GMMLLR}
%
%
In the GW signal detection problem, signal events and noisy background
events are considered as two distinct populations. In this section, we describe
the use of GMMs to model these populations and the development of a detection
statistic based on this. 

\subsection{Log-Likelihood statistic}
\label{sec:LLR}
%

Let us consider the data set with $n$ $d$-dimensional data points as ${\bf X} =
\{{\bf x}_1, {\bf x}_2, \dots{\bf x}_n\}$. We assume that each of the ${\bf
x}_i$'s are independent and we represent ${\bf X}$ as an $n \times d $ matrix.
The likelihood function can then be written as,
\begin{align}\label{maxL}
    p ( {\bf X}| \bm{\theta} ) = \prod _ { i = 1 } ^ { n } p ( {\bf x} _ { i }| \bm{\Theta}), \quad 
\end{align}
where parameters $\Theta := w_j, \boldsymbol{\mu_j}, \Sigma_j, \{j=1,...,K \}$. Then the
corresponding total log-likelihood is the sum of $n$ individual log-likelihoods as given below,
\begin{equation}
    \ln \mathcal{L} =
    \sum _ { i = 1 } ^ { n } \ln ( p ( {\bf x} _ { i }| \bm{\Theta} ) ) 
    = \sum _ { i = 1 } ^ { n } \ln \left\{\sum _ { j = 1 } ^ { K } w_ { j }
\mathcal { N } \left( {\bf x} _ { i }| \boldsymbol{\mu_ j} , \Sigma _ { j }
\right)\right\} \, .\label{eq:like}
\end{equation}

\subsection{Maximum likelihood approach}
%
To estimate the model parameters $\Theta$  we can maximize  $\ln\mathcal{L}$
with respect to each of these parameters. For example,
\begin{align} \label{mu}
\frac { \partial \ln \mathcal{L} } { \partial   \boldsymbol{\mu_ j}}  = 0~~~ \Rightarrow~~~    \boldsymbol{\hat \mu_k} & = \frac { \sum _ { i = 1 } ^ { n } r _ { i k } {\bf x} _ { i }} { N _ { k } }, 
\end{align}
where 
\begin{align}
r _ { i k } &= \frac { w_ { k } \mathcal { N } \left( {\bf x} _ { i }|
\boldsymbol{\mu_k} , \Sigma _ { k } \right) } { \sum _ { j = 1 } w_ { j }
\mathcal { N } \left( {\bf x} _ { i }| \boldsymbol{\mu_j} , \Sigma _ { j }
\right) }\hspace{0.2cm}\text{and}\hspace{0.2cm} N _ { k } = \sum _ { i = 1 } ^ { n } r _ { i k }.
\end{align}
Thus the maximum likelihood estimate of the mean $\boldsymbol{\mu_k}$ of the
$k^{\text{th}}$
Gaussian is the weighted mean of all the data points. All the coefficients are
implicit functions of ${\boldsymbol{\mu_k}}$ {\it via} the normal distributions.

Similarly, maximizing $\ln\mathcal{L}$ with respect to the co-variance matrix
$\Sigma_k$, of the $k^{th}$ Gaussian, we obtain
\begin{align}
    {\hat {\Sigma}} _ { k } = \frac { 1 } { N _ { k } } \sum _ { i = 1 } ^ { n } r _ { i k } \left( {\bf x} _ { i } - {\boldsymbol{\mu_k}} \right) \left( {\bf x} _ { i } - \boldsymbol{\mu_k} \right) ^ { T }.
\label{sigma}
\end{align}

Maximization of Eq.\ref{eq:like} over $w_j$ under the constrain that sum of the
weights add up to unity can be obtained through the application of Lagrange
multipliers. The details can be found in Appendix~\ref{ap1}. This gives the maximum likelihood estimate for weights as
\begin{align}
\hat{w}_ { k } &=  \frac{\sum _ { i = 1 } ^ { n } r_{ik}} {n}  = \frac { N _ {
k } } { n }.\label{maxW}
\end{align}

It is clear from above calculation that it is difficult to analytically
estimate the parameters of the mixture model as all the estimates given in
Eq.~\ref{maxW} are implicit functions of themselves.
 
{The \it{Expectation maximization} (EM)
technique}~\cite{doi:10.1111/j.2517-6161.1977.tb01600.x} is an iterative
algorithm and provides us with a numerical solution to our maximum likelihood  problem.
The EM algorithm has two steps, namely the estimation step and the maximization step. The first
step involves using trial values for the parameters, then, using these values
an iterative step using Eqs.~(\ref{mu}, \ref{sigma}, \ref{maxW}) gives
estimates of the new values of the parameters. Thus, iteratively using
Eqs.~(\ref{mu}, \ref{sigma}, \ref{maxW}), the EM algorithm convergences on the
maximum-likelihood parameters $\hat{\Theta}$.

A drawback of the EM algorithm is that it cannot predict the optimal number of
Gaussians required to describe the underlying structure of the data. Large
numbers of Gaussians can lead to overfitting of the data. The {\it{ Bayesian
information criterion}} (BIC) includes a penalty term which compensates this
effect and is used in the model selection.

The BIC as defined in terms of the maximum value of the likelihood function
$\hat{\mathcal{L}}$ and is given by

\begin{align}
    {\rm BIC} &= K\ln(n) - 2\ln(\hat{\mathcal{L}}).\label{eq:BIC}
\end{align}
where the first term in Eq.~\ref{eq:BIC} is the desired penalty term. The lowest
BIC score provides the optimum number of Gaussians $\hat{K}$ for a given data set.

\subsection{GMM based detection statistic}

Once all the model parameters $\Theta$ are optimally chosen and the optimum number of Gaussians are fixed following the minimum BIC criterion detailed in the previous subsection, we
write the maximum log-likelihood statistic as $W = \ln(\hat{\mathcal{L}} )|_{\hat{K}}$.

Since our data consists of two distinct classes, signals (s) and noisy background glitches (g), we can calculate a detection statistic, $T$, for each trigger such that
\begin{align}
\label{T}
T = W_{\text{s}}- W_{\text{g}}\,.
\end{align}

GMM based models each for the signal and noise are required to determine $W_{\text{s}}$
and $W_{\text{g}}$ respectively. 
In line with tuning and training procedures for cWB and other transient searches, the characterisation and optimisation of the GMM performance is done {\it a prior} on training data, prior to the search for gravitational wave candidates. To calculate $W_{\text{g}}$, we first construct a GMM model using the noise background data. 
The noise background data are divided into a training data set, which is used to construct the GMM model, and a validation data set. 
Similarly, for $W_{\text{s}}$, we construct a GMM model using simulated signals which are also divided into training and validation sets. 
We compute the test statistic $T$ for the validation noise set and validation simulation trigger set and we assess the performance of the GMM models by comparing the detection probability against the false alarm probability. Note that the validation data set allows us to check that the GMM model is overfitting the signal parameter space since it is vital for a generic transient search to be sensitive to a wide range of signal morphologies.

\section{Short duration GW burst search with GMM}
%

%
\label{burst}
A large variety of GW transients fall into the category of short duration
bursts e.g., GWs from supernovae, merger of binary black holes etc. In fact,
the first observed GW signal from a binary black hole merger (GW150914) was
indeed first detected by a generic burst search algorithm coherent Waveburst
(cWB)~\cite{GW150914-DETECTION}. Typically, the cWB burst search method finds
interesting and potential GW events at an initial analysis stage which we refer
as {\it triggers}.  To obtain an estimate of the noise background and assign
each trigger a statistical significance, the cWB analysis is performed on data
that is time-shifted so that it is unphysical for a gravitational wave signal
be detected in coincidence between detectors. The noise background rate is
determined by the number of triggers observed in the time-shifted data.
Triggers observed in data in the absence of an unphysical time shift are
considered {\it event candidates}. These event candidates may be GW signals
though further analysis is often required before any declaration of signal
detection can be made.

To reduce the impact of noise background triggers, thresholds for various
trigger attributes such as its estimated signal strength or duration are
applied. To optimise the chance of detecting a GW signal, these thresholds are
tuned {\it a priori}, using triggers from the time-shifted noise background and simulated
signals. After applying appropriate thresholds on the trigger attributes
pertaining to the search, if a given trigger emerges with high significance
then it is considered a GW event. Such a thresholding procedure, though ad-hoc,
is the crucial ingredient of the signal detection algorithm which helps in
reducing the unwanted noisy features and select the most significant events
(those with very low associated false alarm rate (FAR)) in the data set. 

Here, we introduce GMMs to address this multi-dimensional attribute
thresholding problem where a GMM is constructed for the signal set (and noise
set) defining the characteristic features in the multi-dimentional attribute
space for signal (or noise). Casting this in the log-likelihood based detection
problem as outlined in Sec.~\ref{Sec:GMMLLR} gives a natural map from this
multi-dimentional approach to the scalar log-likelihood ratio. Thus, by
applying a single threshold based on the test statistic across the entire
multi-dimensional attribute space, we can address the ad-hoc individual
attribute thresholding problem in a more systematic way. We henceforth refer to
this signal detection approach as cWB plus GMM.

\subsection{Coherent Waveburst algorithm}
\label{cWB}
For our data sets we consider burst-search triggers from the coherent waveburst
(cWB) algorithm -- a generic multi-detector, all sky burst algorithm used in GW
searches that is sensitive to short duration
signals~\cite{Klimenko:2008fu,klimenko05:_const,Klimenko:2015ypf,
necula12:_trans_wilson_daubec}. This algorithm projects the multi-detector data
into the wavelet (time-frequency map) domain using the
Wilson-Daubechiers-Meyer transformation and identifies a collection of coherent
time-frequency-scale pixels with excess power and clusters them based on the
time-frequency information. Each burst trigger is represented by a coherent
cluster and associated with a set of attributes. The attribute set contains those which characterise 
the signal as well as veto attributes used to help distinguish between signal and noise triggers. The complete set include estimated strain of the trigger $h_s$, the central frequency
$f_0$, the duration of the trigger $\tau$, the network coherent signal-to-noise
ratio $\eta_c$~, the network correlation coefficient $c_c$ which measures
the correlation between the detectors, 
quality factor of the event $Q_{\text{veto}}$, the residual noise energy measure
$\zeta^2$, the ratio between the reconstructed energy and the total energy
$N_{\text{norm}}$, the energy dis-balance of the event between the detectors
$N_{ED}$, and $L_{\text{veto}}$ which measures the localization of event in the
time-frequency map.

The cWB algorithm generates a trigger list that is associated to the
time-frequency clusters with high network correlation $c_c$ as well as cWB SNR
$\eta_c$. Based on the type of the GW signal, the actual threshold values
applied to these two parameters may vary. Following that, the cWB algorithm
uses the additional thresholds on trigger attributes to veto out noisy
triggers. All cWB thresholds are determined by tuning and testing on noise background and simulated signals prior before being used to generate a list of gravitational wave candidates.

\subsection{Data set}
\label{sec:IVB}

To demonstrate our method, we consider cWB triggers obtained from simulated
astrophysical short-duration burst signals added to data from the first advanced
detector observing run O1~\cite{Abbott:2016ezn}. Here, cWB triggers are those
time-frequency clusters with $\eta_c >8.45$ and $c_c > 0.5$.  The 19
types of short duration signals simulated in the O1 data are {\it Gaussian pulses} (GP) characterized by duration $\tau$, {\it sine-Gaussian
wavelets} (SGW) - sinusoids within a Gaussian envelope and characterized by the
frequency$f_0$  and a quality factor $Q$, and {\it White noise bursts} (WNB) -
bursts with a Gaussian envelope described by $f_{\text{low}}$, frequency
bandwidth $\Delta f$, and duration $\tau$. The simulated signal parameter
values used our study is listed in Table~\ref{Tab:BURST}. The signals are
uniformly distributed over the sky and with a range of selected discrete
strain amplitudes from $0.5 \times10^{-21}$-$10^{-20}$. The initial phase is
distributed uniformly over the range $[0,2\pi]$ and the time of arrival of the
signal which depends on the sky location in relation to the detector positions is also distributed uniformly.

The final GW event list is obtained with the standard-cWB after applying
thresholds on the cWB attributes and ranking them based on level of
significance (ascending value of inverse false alarm rate (IFAR)). For more
details on the search see~\cite{Abbott:2016ezn}.

\subsection{Attribute choice}
\label{sec:opt-para}

Each trigger has eleven attributes as defined in Sec.~\ref{sec:IVB}. Here, we
use the exhaustive attribute set $\{c_{c0},c_{c2},\eta_c,f_0, h_s,
N_{\text{norm}},\zeta^2 ,N_{\text{ED}},Q_{\text{veto}},L_{\text{veto}}\}$ to
develop a GMM in the multi-dimentional feature attribute space and use the
minimum BIC method to obtain the optimal attribute set which we use to
construct the GMM for both noise as well as signal data triggers.

For a collection of $n \sim 10^5$ signal triggers from the data set, we
consider the maximum number of Gaussian components to be $100$. We carry out an
exhaustive study to build the signal model. We consider various
combinations of trigger attributes starting from a set of two attributes all
the way to a set of eleven attributes with a varying number of Gaussian
components. We compute the lowest BIC score for a given number of attributes.
For example, there are 55 combinations of pairs of attributes selected from a
set of 11, 165 sets of 3 attributes selected from 11, etc. We construct the BIC
for all possible combinations and choose the minimum value amongst them for a
given attribute count. 

This is done for each attribute count ranging from 2 to 11.

\begin{figure}[ht]
    \centering

    \begin{minipage}[b]{.45\textwidth}
        \includegraphics[scale=0.35]{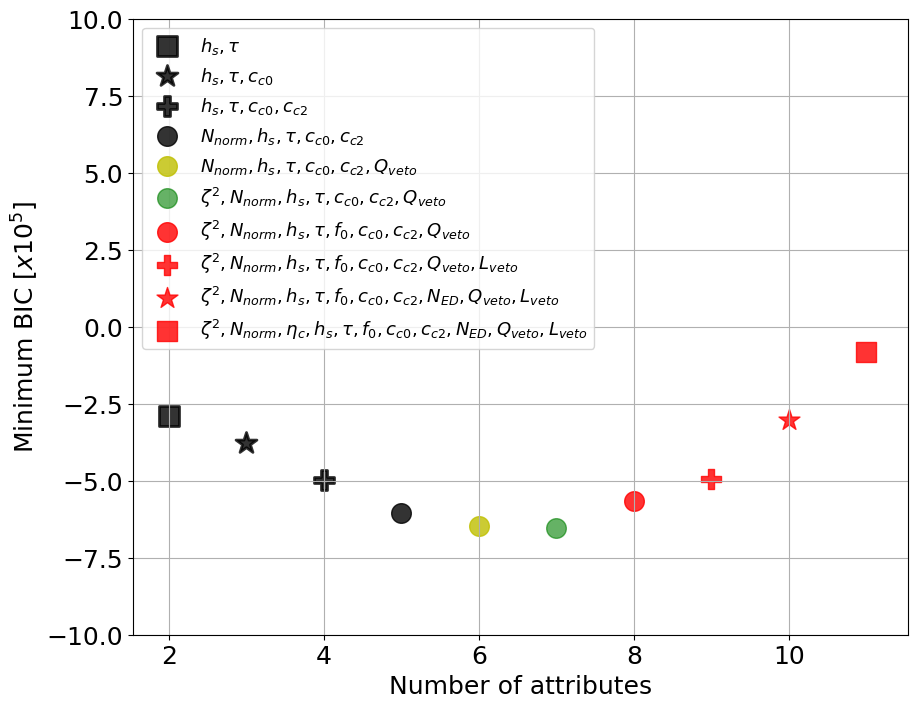}
    \end{minipage}
    \caption{Plot of BIC value {\it{vs}} the number attributes for a given signal data
set. The specific choice of attributes for each set is defined in the
legend.
}
    \label{fig:min_bic}
\end{figure}

\begin{figure*}[!ht]
\includegraphics[scale=0.4]{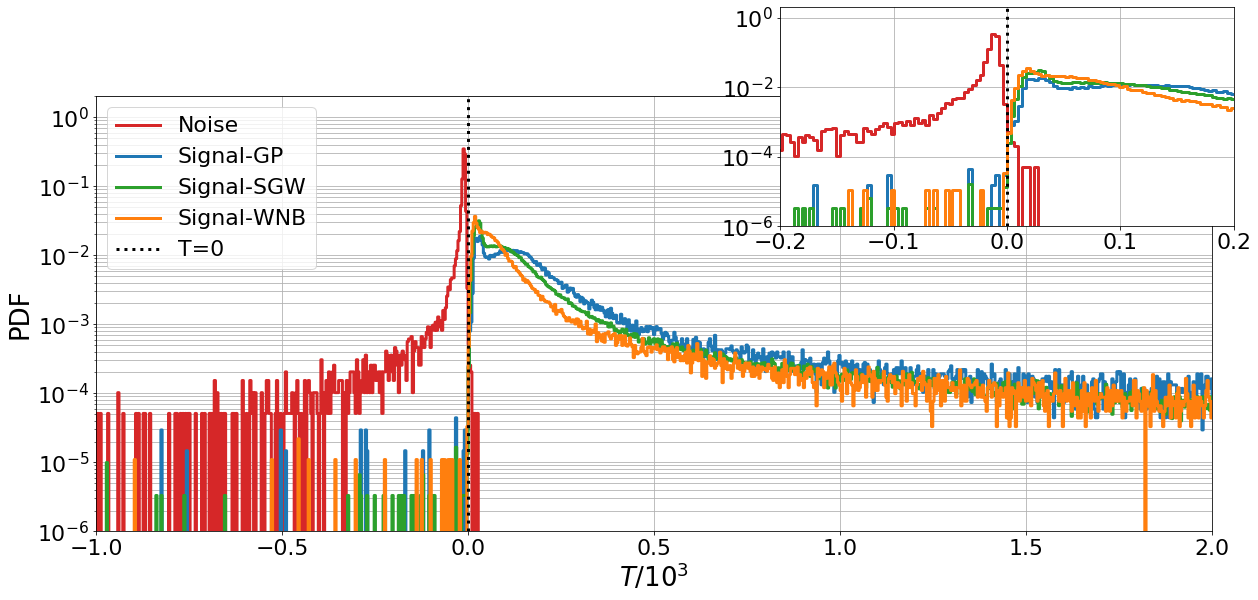}
\caption{The $T$ value distribution for O1 noise and three  different  signal
morphology events with GMM models. Plot contains four curves,  red curve for
noise, blue curve for GP, green curve for SWG and orange curve for WNB. The
inset plot shows the T distribution around zero.}
\label{fig:F2}
\end{figure*}

\begin{figure*}[!ht]
    \hspace{-0.5 cm}
    \centering
    \begin{minipage}[b]{.45\textwidth}
        \includegraphics[scale=0.25]{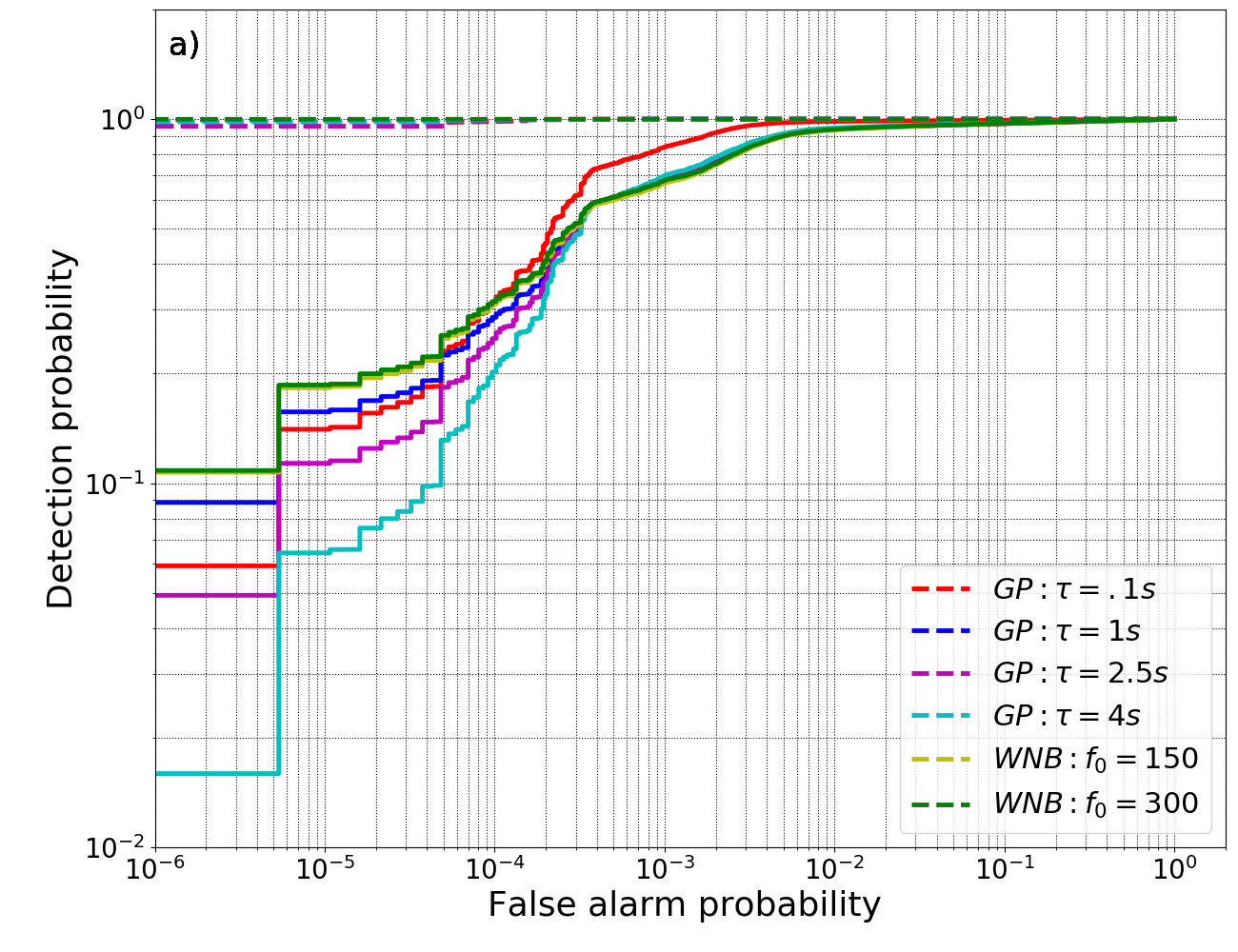}
    \end{minipage}
    \qquad
    \begin{minipage}[b]{.45\textwidth}
        \includegraphics[scale=0.25]{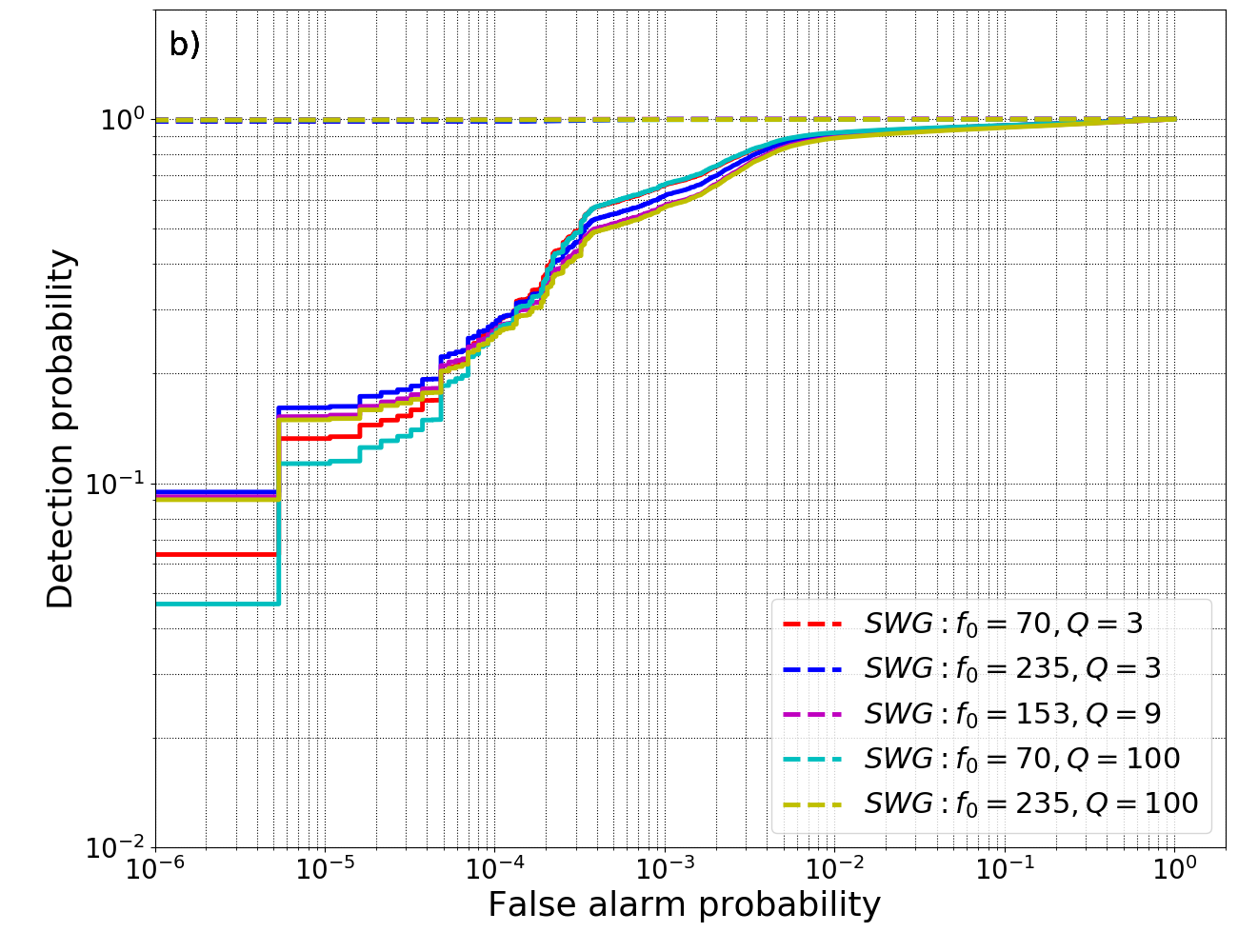}
    \end{minipage}
    \hfill
        \begin{minipage}[b]{.45\textwidth}
        \includegraphics[scale=0.25]{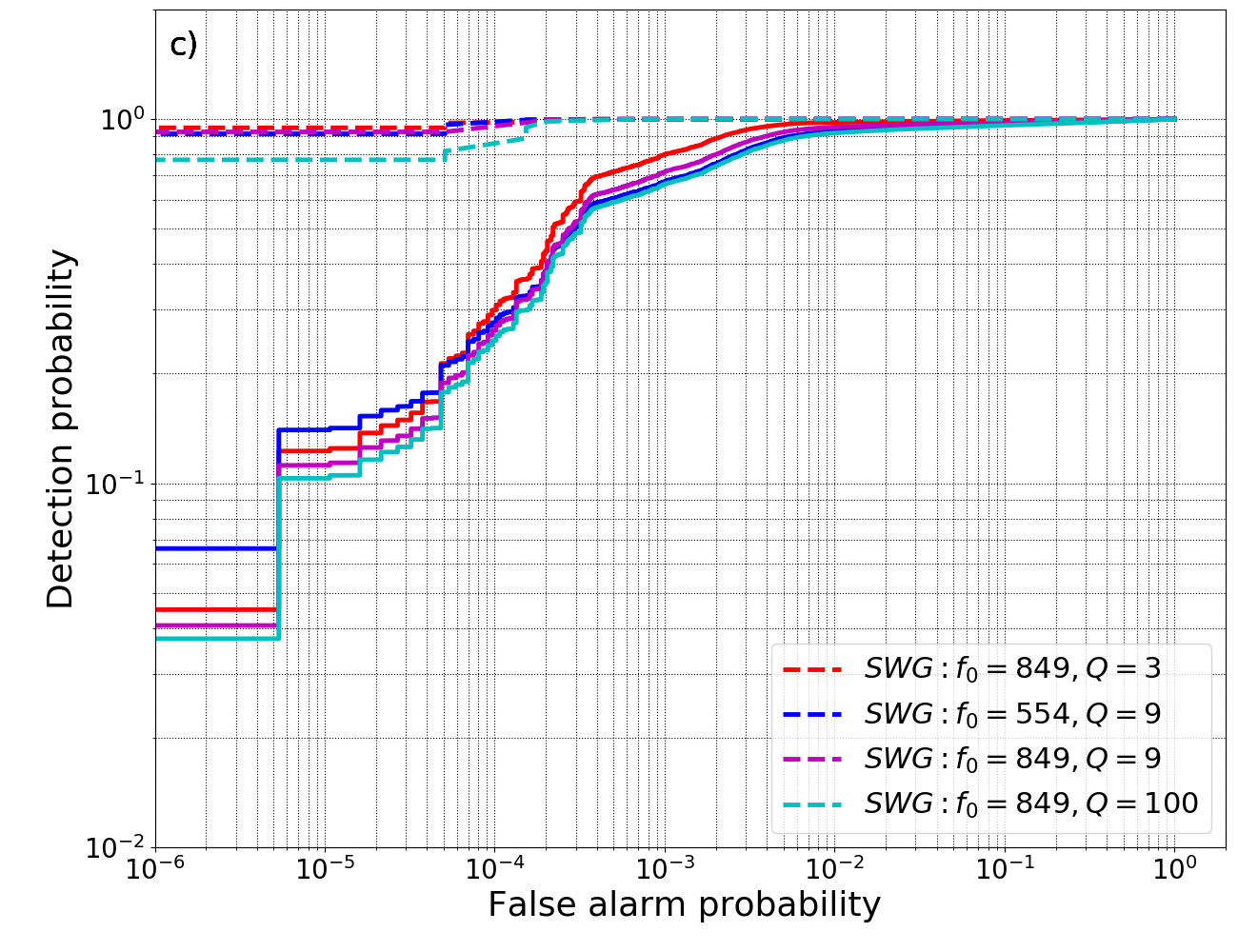}
    \end{minipage}
    \qquad
    \begin{minipage}[b]{.45\textwidth}
        \includegraphics[scale=0.25]{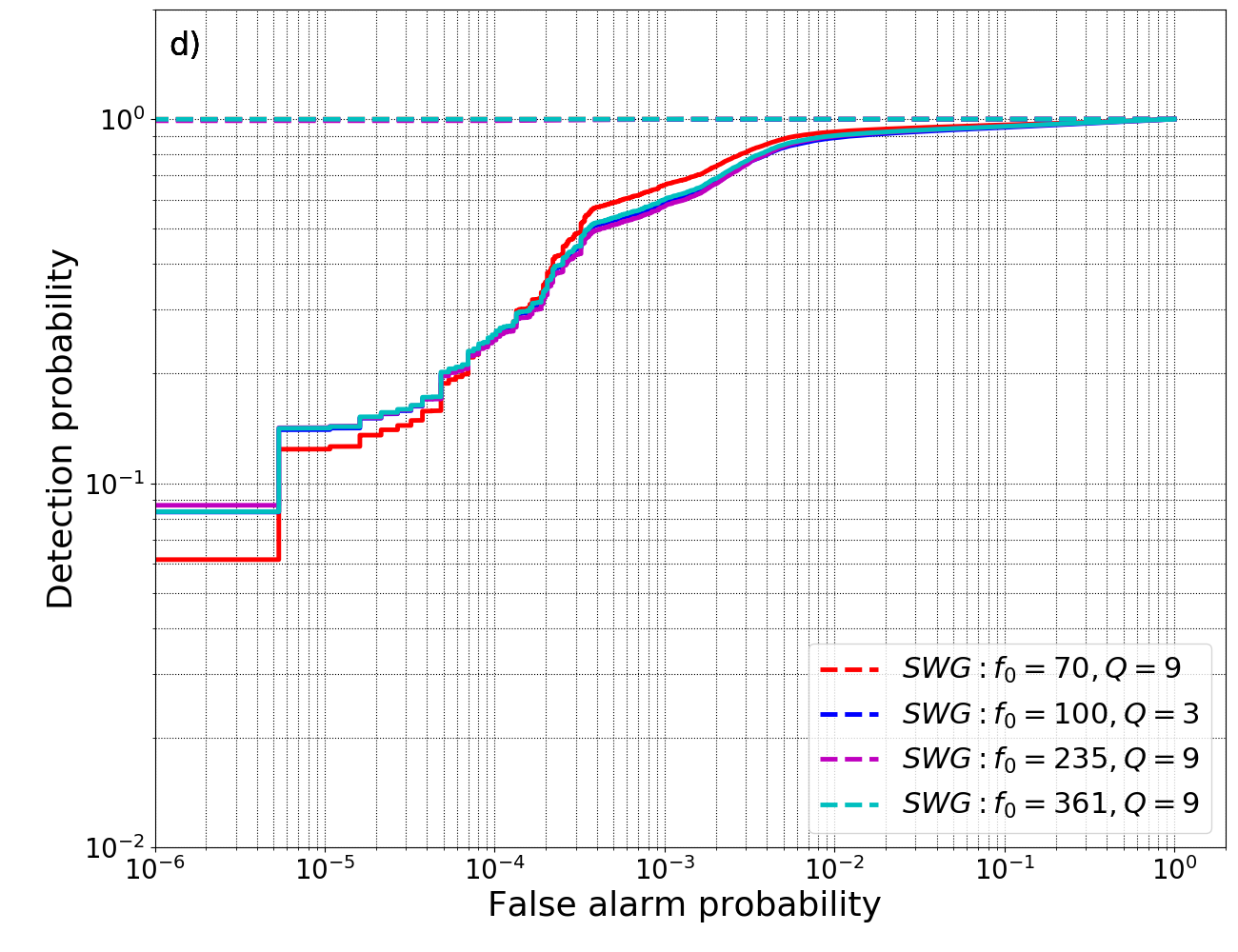}
    \end{minipage}
    \caption{ROC curves: false alarm probability {\it vs} detection probability using the cWB plus GMM algorithm (dashed lines)
compared to the standard-cWB pipeline (solid lines) for the simulation waveforms detailed in Table.\ref{Tab:BURST}. We present the ROC curves in four panels by dividing the simulation sets in different types. For the cWB plus GMM
analysis, the detection probability is close to 0.9 or more for almost all
simulated signals, even at false alarm probabilities of $10^{-6}$.}
\label{fig:ROC}
\end{figure*}

Figure~\ref{fig:min_bic} shows a plot of the minimised BIC value as a function
of the attribute count for the signal model. The combination of attributes
shown in the legend corresponds to the minimum BIC combination for that
attribute count. We note that the minima of this minimum BIC curve corresponds
to the attribute set [$\zeta^2$, $N_{\text{norm}}$, $h_s$, $\tau$, $c_{c0}$,
$c_{c2}$,\footnote{Both $c_{c0}$ and $c_{c2}$ are network correlation coefficients and capture signal correlation between the detectors.} $Q_{\text{veto}}$]-- a seven attribute set which captures the signal triggers. In addition, we note that each set of $p$ optimal attributes contains within it the set of $(p-1)$ optimal attribute set.  
We use the attribute set corresponding to the global BIC minima to construct both the signal and noise trigger
GMMs.

\subsection{Training models}
%

As described in Sec.~\ref{sec:IVB}, the simulated signal injections span a
variety of signal classes that model short bursts with a wide range of
frequencies. 
We aim to develop a GMM for these signals that is robust against the variation
in frequency and duration of the burst. To do so, we develop a minimal model
which broadly captures the burst-like signals with a frequency range of
$70-360$ Hz and different short durations.
Thus, we choose a waveform subset from  the simulation set given in table. \ref{Tab:BURST} which cover the broad class of burst-like signals for model building and test the robustness of this model against the remaining waveform set.  With
the above rationale, our choice is a set of five burst waveforms; SGW with
$f_0=70$Hz \& $Q=3$, SGW with $f_0=235$Hz \& $Q=9$, SGW with $f_0=361$Hz \&
$Q=9$, SGW:$f_0=70$Hz \& $Q=100$ and GP $\tau=1$s as tabulated in the
Table~\ref{Tab:BURST}.

We choose 50\% of the five selected types of burst signal triggers for
training and building the GMM. The remaining signal triggers (comprising all
waveform types) are used for validation, to check that the signal model is robust against varying signal morphologies.
We consider 80\% of noise triggers for
training and building the corresponding noise GMM.
The rest of the noise triggers are used for validation.  This leads to the number
of noise triggers for training being $\sim 10^5$ and signal triggers for training
as $\sim 8 \times 10^4$.

\begin{table}[htb]
\begin{tabular}{c c c c c c}
\hline
\hline
\multicolumn{5}{c}{Sine-Gaussian Burst (SGW)}\\
\hline
No. &$f_0$ (Hz) & $Q$ & - & Training & Validation\\
\hline
1 & 70 & 3 & - & Y & Y\\
2 &70 & 9 & - & N & Y\\
3 &70 & 100 & - & Y & Y\\
4 &100 & 3 & - & N & Y\\
5 &153 & 9 & - & N & Y\\
6 &235 & 3 & - & N & Y\\
7 &235 & 9 & - & Y & Y\\
8 &235 & 100 & - & N & Y\\
9 &361 & 9 & - & Y & Y\\
10 &554 & 9 & - & N & Y\\
11 &849 & 3 & - & N & Y\\
12 &849 & 9 & - & N & Y\\
13 & 849 & 100 & - & N & Y\\
\hline

\multicolumn{5}{c}{White-Noise Burst (WNB)}\\
\hline
 & $f_{low}$ (Hz) & $\Delta f$ (Hz) & $\tau$ (s)& Training & Validation\\
\hline
14 & 100 & 100 & 0.1 & N & Y\\
15 & 300 & 100 & 0.1 & N & Y\\
\hline

\multicolumn{5}{c}{Gaussian Pulse (GP)}\\
\hline
 & - & -& $\tau$ (s) & Training & Validation\\
\hline
16 & - & -& 0.1 & N & Y\\
17 & - & -& 1 & Y & Y\\

18 & - & -& 2.5 & N & Y\\
19 & - & -& 4 & N & Y\\
\hline
\hline
\end{tabular}
\caption{List of generic burst waveforms and their characteristic parameters using in training and validation the GMM signal model.}\label{Tab:BURST}
\end{table}

\subsection{Search sensitivity improvement}
\label{Sec:SensImprove}

In this section, we assess the improvement in search sensitivity by comparing
signal recovery with the standard-cWB algorithm and the cWB plus GMM based
detection algorithm.

After constructing the GMMs for signal and for noise triggers,
Fig.~\ref{fig:F2} shows the distribution of the detection statistic $T$ for
noise and signal testing data computed using Eq.~\ref{T}. The plot shows four
curves, noise (red), GP signals (blue), SWG signals (green), and WNB signals
(orange). We observe that the majority of noise triggers have negative $T$
value and similarly most signal triggers have positive $T$ value. There is a
corresponding clear separation between the signal and the noise triggers at the
$T=0$ boundary. Very small overlap exists between the T distributions for
signal and noise triggers. This clearly indicates that the GMM  based proposed
likelihood ratio statistic has the potential to discriminate between the two
classes of triggers. 
   
To compare the performance between the standard-cWB and the cWB plus GMM based
detection approach, we construct Receiver Operating Characteristic (ROC)
curves. ROC curves relate the detection probability and the false alarm
probability as the threshold on the test statistic is changed. For the
standard-cWB analysis, this means recording the fraction of simulated signals
and noise triggers for varying threshold values on $\eta_c$, while for cWB plus
GMM, it is the threshold on $T$ that is varied. We use $\sim 3 \times 10^4$
noise triggers and $10^5$ signal triggers to create the ROC curves shown in
Fig.~\ref{fig:ROC}.

We note that cWB plus GMM based detection algorithm significantly enhances the
signal detection probability when compared to the standard-cWB detection alone.
The detection probability using the cWB plus GMM varies between 0.9 and 1 for
most of the waveform models. The performance for SWG waveform with
$f_0=849$Hz and $Q=100$.  (see panel c in Fig.~\ref{fig:ROC}) appears to have dropped a
little compared to other waveforms. This is primarily because the cWB plus GMM
approach is sensitive to the parameter distributions used within the training
data and we have chosen to use burst waveforms with frequency values up to 350
Hz and short durations. For high frequency signals, although we do not get as spectacular
performance as the other cases, we do achieve significantly improved performance
compared to the standard-cWB settings.

\subsection{Application of cWB plus GMM to the O1 data}
\label{Sec:AppToO1}
 
We apply the cWB plus GMM algorithm to the coincident data from O1. The GMM
signal model is already trained using the all-sky short duration cWB triggers obtained
from simulations as detailed in Sec.\ref{Sec:SensImprove}. The GMM noise model is also
trained on the time-shifted cWB background triggers from the O1 all-sky short
duration search. We compare the results after applying the cWB plus GMM
algorithm with the standard-cWB results obtained by the all sky short duration
burst search reported in \cite{Abbott:2016ezn}.

\begin{figure}[t]
\includegraphics[scale=0.4]{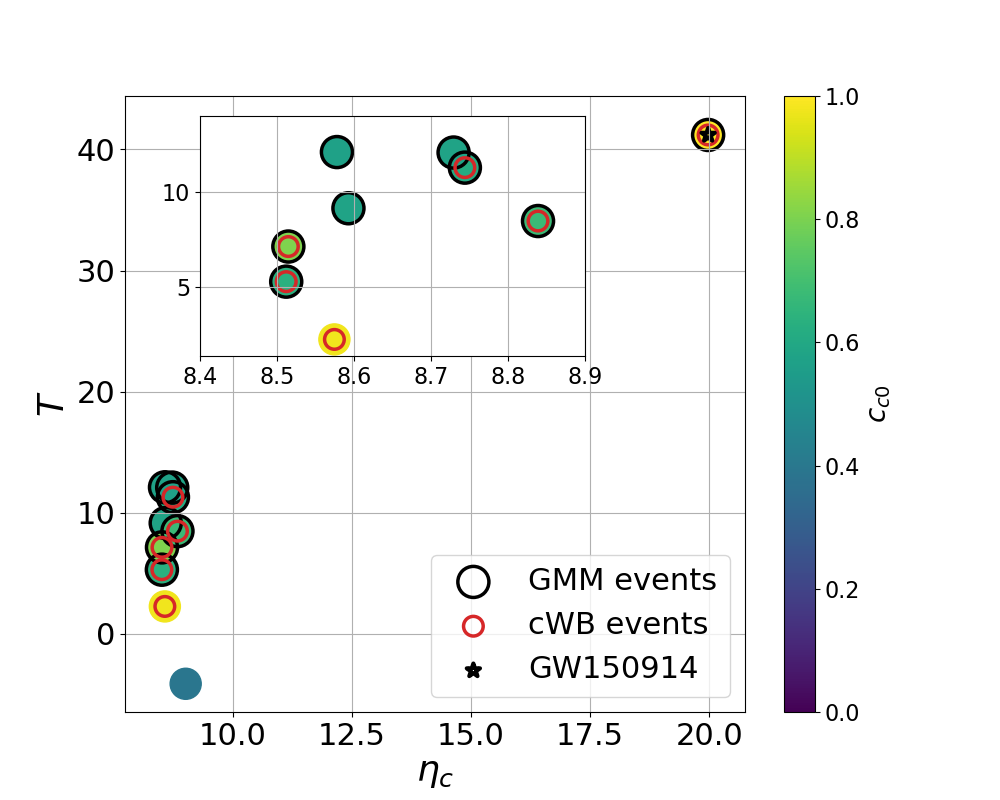}
\caption{The network coherent SNR $\eta_c$ vs the GMM detection statistic $T$,
with color corresponding to $c_{c0}$ for all 10 cWB triggers.  
Large black hollow circles correspond to detected cWB with GMM events (with
$T>2.5$ equivelent to IFAR$>15$years) and smaller red hollow circles correspond to the
standard-cWB events. The black star corresponds to the GW150914 event}
\label{fig:zero} 
\end{figure}

For the O1 all-sky short duration search, a total of 10 cWB triggers are
observed when no time shift is applied to the data.  For the standard-cWB
algorithm, additional thresholds are applied on the cWB trigger attributes to
further reject the noise triggers. The additional threshold requirements used
in the standard-cWB used for the O1 all-sky short duration search are $c_{c0} >
0.7$, $48$Hz$<f_0<998$Hz, $Q_{\text{veto}}>0.3$ and $\zeta^2 <0.5$. A total of
6 coincident events survive these threshold criteria~\cite{Abbott:2016ezn}. We
draw these events as red hollow circles in Fig.~\ref{fig:zero}.

Instead of applying various thresholds on individual trigger attributes, we
consider the cWB plus GMM based detection approach on the triggers. We compute
the $T$ statistic for each cWB trigger and select events with $T>2.5$
(corresponding to an IFAR of 15 years) as candidate events denoted by
large black hollow circles in Fig.~\ref{fig:zero}. For the IFAR estimation, we consider the entire background trigger set. 

Figure~\ref{fig:zero} shows $\eta_c$ vs $T$, with color corresponding to
$c_{c0}$ for all cWB triggers ($\eta_c>8.45$ and $c_{c0}>0.5$). The inset plot
shows $\eta_c$ vs $T$ for the events clustered below $\eta_c=8.9$. We show all
10 cWB triggers as filled circles with the colour indicating the value of
$c_{c0}$. We notice that 5 events are common in both the searches. In addition,
the cWB plus GMM identifies three events which were rejected by the
standard-cWB analysis and one event from standard-cWB is rejected by cWB plus
GMM.

Both standard-cWB and cWB plus GMM observe GW150914 to be the most significant
event, which is indicated by a star in Fig.~\ref{fig:zero}. For cWB plus GMM,
GW150914 is observed with $T\sim41$. With no background events having that $T$
value or greater, the cWB plus GMM analysis detects GW150914 with an IFAR of
greater than $1000$ years. We note that the standard-cWB analysis for O1
detects GW150914 with an IFAR of $\sim 350$ years, even after the analysis of
O1 data was split into 3 classes to isolate chirp-like signals into a single
class to separate them from spurious noise transients~\cite{Abbott:2016ezn}. The remaining four common events also show higher significance in cWB plus GMM
compared to the standard-cWB. The two additional events observed by the cWB
plus GMM have network correlation just below 0.7 and hence went undetected by
the standard cWB.

\section{Conclusion}
\label{conclusion}

Detection of transient GWs in non-stationary and non-Gaussian noise poses a
massive challenge in the advanced laser interferometric gravitational wave
detectors. A number of approaches are used to combat spurious noisy triggers
and to allow the detection of astrophysical GW transients with high
significance.  The standard-cWB GW burst search {\it first} makes a list of
triggers -- interesting time instances which could be either potential GW
signals or background noise transients -- and {\it then} applies a list of
thresholds on the attributes to reject the noisy triggers.  Although this
thresholding approach is guided by extensive simulations and hence is largely
successful, it is still ad-hoc in nature and relies on analysts intuition to
optimise the thresholding process. In this work, we propose a supervised
machine learning method using GMMs trained on cWB triggers.  This training is
applied using the O1 all-sky search triggers from the cWB search for short
duration GW bursts to develop two distinct models for noise triggers and
astrophysical GW burst triggers. We use these models to construct a
log-likelihood based test statistic. We demonstrate that this approach gives
improved performance as compared to the standard-cWB approach by improved
signal detection probability at any FAR. We also obtain a significantly
improved detection significance for GW150914 (the first ever GW merger event)
with the cWB plus GMM detection approach. With this example, we clearly
demonstrate that our more systematic GMM based signal detection approach can
improve the detection performance as compared to the thresholding approach in
the multi-dimensional attribute space.

It is worth noting that, like most GW search algorithms, the GMM approach is
sensitive to simulation parameters used to train the signal model. Here, for
comparison with standard-cWB, we have used the population of simulated signals
that were generated for the O1 all-sky short duration burst search. The
simulated data set was designed to provide an estimate of the search
sensitivity to an ad-hoc population of signals and was not drawn from an
astrophysical population. For example, the simulated data did not uniformly
span the frequency band of the detectors. If we had included frequency as one
of the parameters for the GMM, then the resulting test statistic would favour
signals with frequencies corresponding to the simulated population which is not
a desirable feature for unmodelled burst searches.

By applying the GMM to cWB trigger attributes, we are relying on cWB to
efficiently capture all characteristics of an interesting event in these
attributes. The cWB algorithm has undergone almost two decades of development
and much effort has been put into the development and characterisation of the
algorithm. Nonetheless, it may be possible to find more efficient ways to
encode GW data for more effective GMM signal and noise model construction. For
example, neural networks can be used to map GW strain data into a reduced
number of parameters which are then used as inputs into the corresponding GMM.

In the near term, we plan to test this GMM approach using data from the second
advanced detector observing run which contains a large variety of noisy
triggers. The method is general enough and is not limited to short duration
bursts triggers. We plan to extend this approach to GW searches for
intermediate mass binary black hole mergers as such signals can be very
difficult to distinguish from spurious noise transients, especially with
increasing black hole mass.

\section{Acknowledgements}
The authors thank F. Salemi for useful comments. VG acknowledges Inspire division, DST, Government
of India for the fellowship support. VG acknowledges support from the University of Florida. DL and RP thank Newton-Bhabha fund for the travel
support to visit the University of Glasgow. AP acknowledges support from the SERB
Matrics grant MTR/2019/001096 as well as the IIT-Bombay SEED grant for the
travel funds to host the visit of ISH and CM to IIT Bombay. CM and ISH are
supported by the Science and Technology Research Council (grant No. ST/
L000946/1) and the European Cooperation in Science and Technology (COST) action
CA17137.

\appendix

\section{Analytical Maximization of log likelihood}
\label{ap1}

Here, we analytically maximize the log-likelihood function given in
Eq.~\ref{eq:like} under the constraint of $\sum_{j=1}^{K}w_{j} =1$. To estimate the weights $w_{1},w_{2},\dots,w_{K}$, we apply
the method of Lagrange multipliers.  Thus, we maximize

\begin{align}
\mathcal{L}^{\prime}&=\ln(p(X|\bm{\Theta}))+\lambda\left(\sum_{j=1}^{K}w_{j}-1\right)
\end{align}
with respect to $w_k$. This gives,
\begin{align} \label{eq:est_phi}
    \frac { \partial \mathcal{L} ^ { \prime } } { \partial w _ { k } } = \sum _ { i = 1 } ^ { n } \frac { \mathcal { N } \left( \vec{x} _ { i } | \boldsymbol{\mu_k} , \Sigma _ { k } \right) } { \sum _ { j = 1 } ^ { K } w _ { j } \mathcal { N } \left( \vec{x} _ { i } | \boldsymbol{\mu_k } , \Sigma _ { k } \right) } + \lambda = 0
\end{align}
Multiplying the above equation by $w_k$ and summing over $k$ we get,
\begin{align*}
 \sum _ { i = 1 } ^ { n } \frac {\sum _ { k = 1 } ^ { K } w_k \mathcal { N } \left( \vec{x} _ { i } | \boldsymbol{\mu_k} , \Sigma _ { k } \right) } { \sum _ { j = 1 } ^ { K } w _ { j } \mathcal { N } \left( \vec{x} _ { i } | \boldsymbol{\mu_ k} , \Sigma _ { k } \right) } + \lambda\sum _ { k = 1 } ^ { K } w_k = 0.
\end{align*}
Summing over $i$ we obtain $\lambda=-n$. Substituting $\lambda$ in Eq.~(\ref{eq:est_phi}) and multiplying by $w_k$ we obtain, 
\begin{align}
n w_ { k } &=  \sum _ { i = 1 } ^ { n } r_{ik}  \Rightarrow  w_ { k } = \frac { N _ { k } } { n } 
\end{align}

Therefore, the weight $w_k$ for the $k^{\text{th}}$ Gaussian
component is given by the the average contribution of $N_k$.

\bibliography{reference.bib}
\end{document}